\documentstyle[epsf,twocolumn,seceq]{orb2001}

\title
{
Electronic Structures of Antiperovskite Superconductor
MgCNi$_3$ and Related Compounds
}

\def\runtitle
{
Electronic structures of antiperovskite superconductor 
MgCNi$_3$ and related compounds
}

\author
{
J. H. {\sc Shim},\footnote{e-mail: jhshim@postech.ac.kr}
S. K. {\sc Kwon} and
B. I. {\sc Min}
}

\def\runauthor
{
J. H. {\sc Shim} {\it et al.}
}

\inst
{
 Department of Physics, Pohang University of Science and
Technology, Pohang 790-784, Korea
}

\abst
{
Electronic structure of a newly discovered antiperovskite 
superconductor MgCNi$_3$ is investigated by using the 
LMTO band method.
The main contribution to the density of states (DOS) at the
Fermi energy $E_{\rm F}$ comes from Ni 3$d$ states which are 
hybridized with C 2$p$ states. The DOS at $E_{\rm F}$ is varied
substantially by the hole or electron doping due to
the very high and narrow DOS peak located just below $E_{\rm F}$.  
We have also explored electronic structures of C-site and Mg-site 
doped MgCNi$_3$ systems, and described the superconductivity
in terms of the conventional phonon mechanism. 
}

\kword
{
Antiperovskite, Superconductor,  Electronic structure, MgCNi$_3$
}

\begin{document}
\sloppy
\maketitle

\section{Introduction}

In contrast to the ordinary perovskite structure,
antiperovskite structure has metal elements at the corners of
the octahedron cage. Ordinarily transition metal based antiperovskites 
have magnetic ground states. For example, Mn-based antiperovskites have
been studied due to their complex magnetic structure\cite{fruchart}
and large magnetoresistance\cite{kamishima,kim}.
Recently, superconductivity is observed in antiperovskite MgCNi$_3$
with transition temperature of $T_c \sim$ 8 K by He {\it et al.}\cite{he}.
It is rather suprising that the superconductivity is observed in such
materials with much Ni contents. Band calculations show that 
very narrow and high peak is located just below $E_{\rm F}$ in density of
states (DOS) curve\cite{hayward,shim,dugdale,singh,rosner}. 
This peak
has mainly Ni 3$d$ character hybridized with C 2$p$ states\cite{shim}.

Superconductivity of MgC$_x$Ni$_3$ is very sensitive to the
C contents $x$. It disappear inbetween $x =$ 0.96 $\sim$ 0.90\cite{he}.
By Ni-site doping, superconductivity is also suppressed either
hole with Co or electron doping with Cu, respectively\cite{hayward}.
Especially, Co doping do not induce magnetism\cite{szajek},
and so the magnetism is not the origin of suppression of the
superconductivity. Hence the mechanism of
abrupt quenching of superconductivity is not known yet.
It is also expected that Mg-site hole doping will produce the magnetic 
ground state\cite{singh,rosner}.

In this work, we have explored C-site
and Mg-site doping effects with substituting neighboring atoms 
in the periodic table.  By comparing the DOS, Stoner factor, 
and electron phonon coupling constant, we have discussed the effect 
of doping on the superconductivity and magnetism.

\section{Computational details}

We have used the linearized muffin-tin orbital (LMTO) band method in the
local density approximation (LDA). The LMTO band calculation 
include muffin-tin orbitals upto $d$-site for Mg, C, and upto $f$-site for Ni.
We have considered a cubic structure with lattice constant
$a = 3.816$ \AA, which is for the highly stoichiometric MgCNi$_3$ with
highest T$_c$ \cite{he}.
Atomic positions are Mg at (0 0 0), C at (1/2 1/2 1/2), and Ni at
(1/2 1/2 0), (1/2 0 1/2), and (0 1/2 1/2).
Atomic sphere radii of Mg, C, and Ni are employed as 3.20, 1.54
and 2.49 \AA, respectively.  For the calculation of doped MgCNi$_3$ systems, 
we have considered a supercell with the doubled unit cell.
One type of Mg or C (111) planes alternate with other type planes 
along the [111] direction. 
Without the knowledge of real structural information of doped 
systems, we have assumed the same structural parameters.

To check the magnetic instability, We calculate the Stoner exchange enhancement 
parameter S, defined as $S\equiv N(E_{\rm F})I_{\rm XC}$ with $I_{\rm XC}$ 
denoting the intraatomic exchange-correlation integral.
To investigate the superconducting properties of MgCNi$_3$ based on
the rigid-ion approximation\cite{gaspari}, the 
electron-phonon coupling constant
$\lambda_{\rm ph}$ is evaluated by McMillan's formula\cite{mcmillan}
\begin{equation}
      \lambda_{\rm ph} = \sum_{\alpha}
      \frac{N(E_{\rm F})\langle I^2_{\alpha}\rangle}
      {M_{\alpha}\langle\omega_{\alpha}^{2}\rangle},
\end{equation}
where $\langle I^2_{\alpha}\rangle$ is the average electron-ion interaction
matrix element for the $\alpha$-th ion, $M_\alpha$ is an ionic mass and
$\langle\omega_{\alpha}^{2}\rangle$ is the relevant phonon frequency.
Since we have no information about phonon frequency, we instead use
the average phonon frequency $\langle\omega^{2}\rangle\simeq\Theta_{D}^{2}/2$,
where $\Theta_{D}$ is the Debye temperature. $\Theta_D$ is adopted for 300 K
and 400 K.

\section{Electronic structure of MgCNi$_3$}

\begin{figure}[t]
  \vspace{10pt} 
     \epsfxsize=8cm 
     \epsfigure{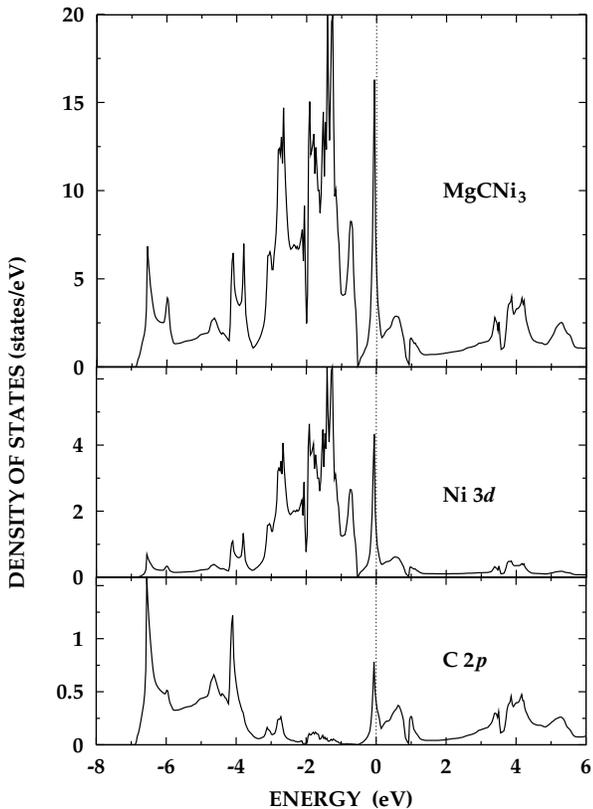}
  \caption{The total and partially projected local DOS of MgCNi$_3$}
  \label{dos}
\end{figure}

Total and partially projected local DOSs of MgCNi$_3$ are presented
in Fig.\ \ref{dos}.
It is seen that Ni 3$d$ and C 2$p$ states are strongly
hybridized. The peak near $-$7 eV and 4 eV correspond to $\sigma$ bonding
and $\sigma^*$ antibonding states, respectively. On the other hand, 
the peak near $-$4 eV and $E_{\rm F}$ correspond to $\pi$ bonding
and $\pi^*$ antibonding states, respectively.
Because of the  $\pi^*$ antibonding states near $E_{\rm F}$, 
the DOS is mainly composed
by Ni 3$d$ and C 2$p$ states as shown in Table\ \ref{dos-table}.
Especially, the contribution of Ni 3$d$ states to the N($E_{\rm F}$)
is as much as 76\%. 
Due to very narrow and high peak in DOS located just below
$E_{\rm F}$, the system is expected to be perturbed much
by small hole or electron doping. The Stoner parameter of S = 0.64 shows
that this system is not magnetic, but it is expected to become
magnetic by small hole doping in the rigid band scheme. 

With the choice of $\Theta_D =$ 400 K, one obtains 
$\lambda_{\rm ph}$ = 0.77. McMillan's $T_c$ formula with an effective
electron-electron interaction parameter $\mu^{*}$ = 0.13 gives rise to
$T_c$ = 11 K, which are in reasonably agreement with experiment.
The superconducting property will be also rapidly changed by
hole or electron doping.

\begin{table}[b]
\vspace{12pt}
\caption{The calculated density of states at $E_{\rm F}$ (in 
states/eV) for MgCNi$_3$.}
\label{dos-table}
\begin{center}
\begin{tabular}{@{\hspace{\tabcolsep}\extracolsep{\fill}}cccccc}
\hline
 & $s$ & $p$ & $d$ & $f$ & Total  \cr
\hline
 Mg  & 0.00   & 0.19   & 0.03   &      & 0.22  \cr
 C   & 0.02   & 0.40   & 0.00   &      & 0.42  \cr
 Ni  & 0.07   & 0.13   & 1.37   & 0.01 & 1.58  \cr
\hline
\end{tabular}
\end {center}
\end{table}

\section{Electronic structure of C-site and Mg-site doped MgCNi$_3$}

\begin{figure}[t]
  \vspace{10pt}  
     \epsfxsize=8cm 
     \epsfigure{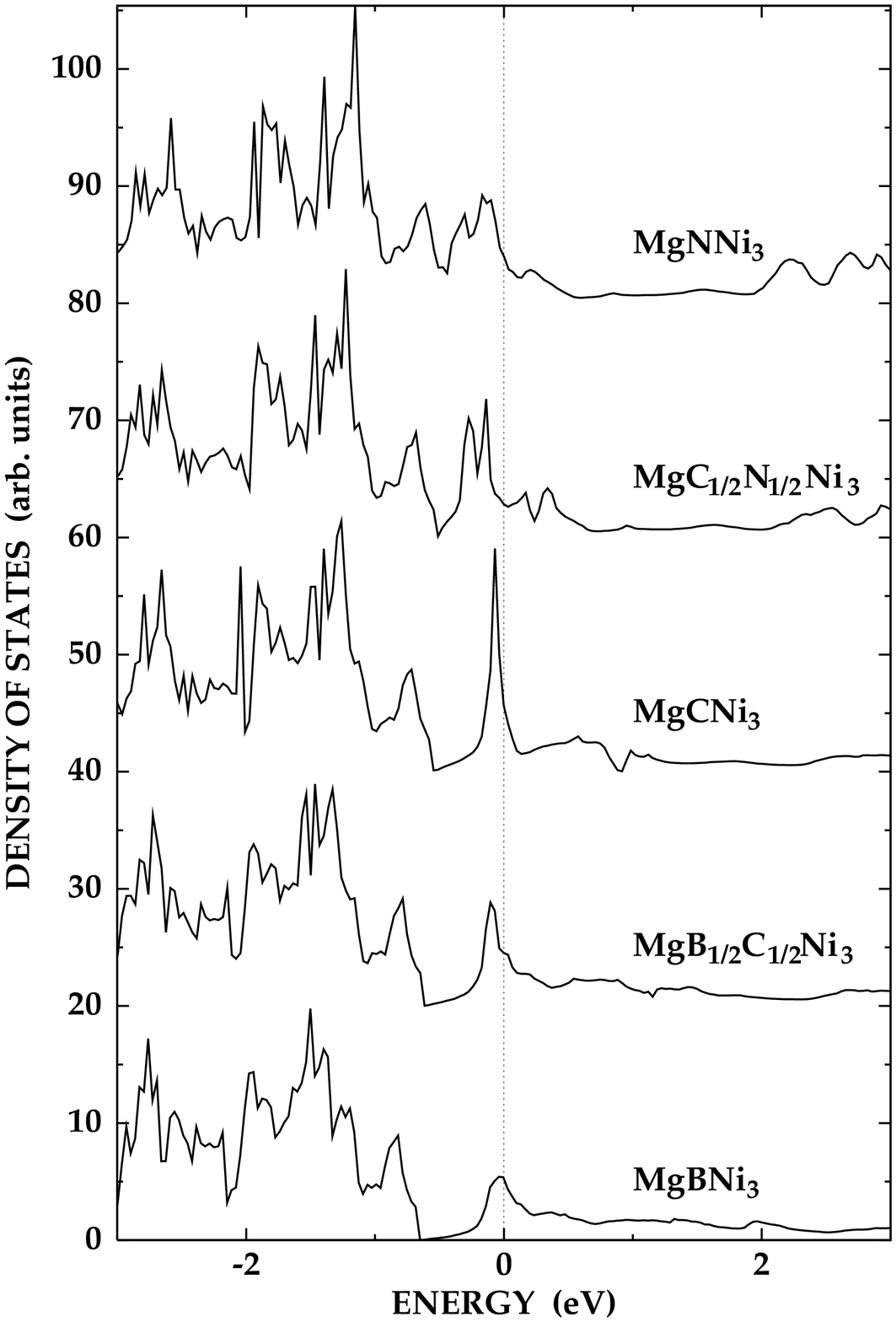}
  \caption{Comparison of total DOSs by C-site doping}
  \label{cdoping}
\end{figure}
\begin{table}[b]
\vspace{12pt} 
\caption{Comparison of total DOS at $E_{\rm F}$, N($E_{\rm F}$) (in states/eV), 
Stoner factor S, and electron phonon coupling constant $\lambda_{\rm ph}$ for
$\Theta_D = $ 300 K and 400 K by C-site doping.}
\label{cdoping-table}
\begin{center}
\begin{tabular}{@{\hspace{\tabcolsep}\extracolsep{\fill}}lcccc}
\hline
 & $ N(E_{\rm F})$            & S 
 & $ \lambda_{\rm ph}$(300K)  & $\lambda_{\rm ph}$(400K) \cr
\hline
$\rm MgBNi_3$               & 5.26   & 0.60   & 0.44   & 0.25  \cr
$\rm MgB_{1/2}C_{1/2}Ni_3$  & 4.61   & 0.53   & 0.78   & 1.11  \cr
$\rm MgCNi_3$               & 5.56   & 0.67   & 1.36   & 0.76  \cr
$\rm MgC_{1/2}N_{1/2}Ni_3$  & 2.84   & 0.34   & 0.59   & 0.33  \cr
$\rm MgNNi_3$               & 4.01   & 0.51   & 1.16   & 0.65  \cr
\hline
\end{tabular}
\end {center}
\end{table}
\begin{figure}[t]
  \vspace{10pt}
     \epsfxsize=8cm 
     \epsfigure{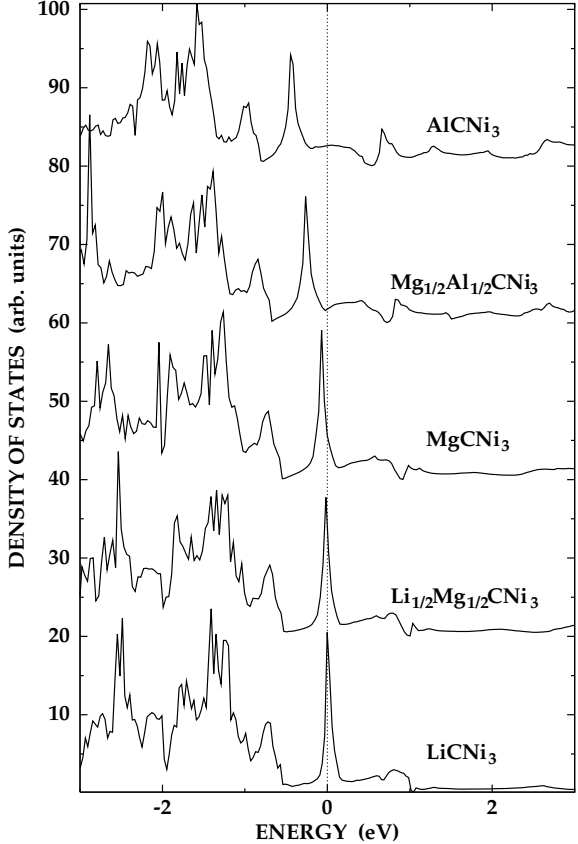}
  \caption{Comparison of total DOSs by Mg-site doping}
  \label{mgdoping}
\end{figure}
To see the doping effect in a rigid band way,
Mg-site or C-site doping would be more adequite rather
than Ni-site doping.  Ni-site doping would perturb
electronic structure near $E_{\rm F}$ substantially
due to the high contribution of doped element.
We have examined quite a few doped MgCNi$_3$ systems
to see the doping effect systemically.

Firstly, C-site doping is studied by doping with B and N, which
correspond to one hole and electron doping, respectively. 
As shown in Fig.\ \ref{cdoping}, the effect of doping is mainly
a variance of Fermi level, even though the shape of DOS near 
$E_{\rm F}$ is perturbed a little bit.
Doping with B smears the DOS peak near $E_{\rm F}$, because
the upward shift of B 2$p$ state induce the strong hybridization with
Ni 3$d$. Although the Fermi level MgCNi$_3$ is just located at DOS peak,
The DOS at $E_{\rm F}$ is not so high.
By doping with N,
the width of the DOS peak between $E_{\rm F}$ and 1.0 eV in MgCNi$_3$ 
is reduced so that the DOS shape near $E_{\rm F}$
is more perturbed. The DOS is not described by a systematic upward
shift of the Fermi level. 

Total DOS at $E_{\rm F}$, N($E_{\rm F}$), the Stoner parameter S, 
and the electron-phonon coupling constant $\lambda_{\rm ph}$ are compared
for several doped systems in Table\ \ref{cdoping-table}. 
Contrary to the expectation in the  rigid band scheme,
N($E_{\rm F}$) is highest in MgCNi$_3$ and do not show the systematic
behavior. For all cases, the magnetic instability is not shown.
From $\lambda_{\rm ph}$, it is expected that electron or hole doping
would suppress the superconductivity. But, MgNNi$_3$ is expected to
have comparable $T_c$ to MgCNi$_3$, once synthesized successfully
in the antiperovskite structure.
\begin{table}[b]
\vspace{12pt}
\caption{Comparison of total DOS at $E_{\rm F}$, N($E_{\rm F}$) (in states/eV),
Stoner factor S, and electron phonon coupling constant $\lambda_{\rm ph}$ for
$\Theta_D = $ 300 K and 400 K by Mg-site doping.}
\label{mgdoping-table}
\begin{center}
\begin{tabular}{@{\hspace{\tabcolsep}\extracolsep{\fill}}lcccc}
\hline
 & $ N(E_{\rm F}) $           & S 
 & $ \lambda_{\rm ph}$(300K)  & $\lambda_{\rm ph}$(400K) \cr
\hline
$\rm LiCNi_3$                & 20.43   & 2.43   & 2.86   & 1.61  \cr
$\rm Li_{1/2}Mg_{1/2}CNi_3$  & 15.27   & 1.87   & 2.15   & 1.21  \cr
$\rm MgCNi_3$                & 5.56    & 0.67   & 1.36   & 0.76  \cr
$\rm Mg_{1/2}Al_{1/2}CNi_3$  & 1.81    & 0.21   & 0.39   & 0.22  \cr
$\rm AlCNi_3$                & 2.61    & 0.29   & 0.52   & 0.29  \cr
\hline
\end{tabular}
\end {center}
\end{table}

Mg-site doping is expected to be better described in the rigid band scheme,
because the contribution of Mg to N($E_{\rm F}$) is negligible
in MgCNi$_3$. We have performed the Mg-site doping with Li and Al,
which have one less and one more electron. We have chosen Li as
an effective hole doping element rather than Na, because the rigid
band scheme works much better. As shown in Fig.\ \ref{mgdoping},
the doping results in only the shift of the Fermi level. The shape
of DOS near $E_{\rm F}$ is well maintained.

Table\ \ref{mgdoping-table} shows a systematic behavior of N($E_{\rm F}$)
by doping. By hole doping, N($E_{\rm F}$) becomes much larger 
to produce the magnetic instability.
Both LiCNi$_3$ and Li$_{1/2}$Mg$_{1/2}$CNi$_3$ have ferromagnetic ground
states with magnetic moments of 0.74 $\mu_B$ and 0.82 $\mu_B$.
Therefore, Mg-site hole doping is expected to induce the transition from
superconducting to magnetic instability.
On the other hand, in the case of electron doping, 
N($E_{\rm F}$) becomes much smaller than in MgCNi$_3$, 
resulting in the suppression of superconductivity.

\section{Conclusion}

We have investigated the electronic structure and 
superconducting property of MgCNi$_3$.
The $\pi^{*}$ antibonding state of Ni 3$d$ and C 2$p$ states
is located just below $E_{\rm F}$. To explore the doping effect,
we have considered the effective hole and electron dopings:
at C-site by B and N, and at Mg-site by Li and Al. 
Mg-site doping is described better by the rigid band doping effect 
rather than C-site doping.
Mg-site doping with Li is expected to show the transition from  
superconducting to magnetic ground state.

\section*{Acknowledgements}

This work was supported by the KOSEF through the eSSC at POSTECH
and in part by the BK21 Project. Helpful discussions with N.\ H.\ Hur are
greatly appreciated.


\end{document}